\documentclass[12pt]{article}
\pdfoutput=1
\usepackage{epsfig}
\usepackage{amsfonts}
\usepackage{amssymb}
\usepackage{xcolor}

\begin{document}
\newcommand{\nc}{\newcommand}
\nc{\beq}{\begin{equation}} \nc{\eeq}{\end{equation}}
\nc{\beqa}{\begin{eqnarray}} \nc{\eeqa}{\end{eqnarray}}
\nc{\R}{{\cal R}}
\nc{\A}{{\cal A}}
\nc{\K}{{\cal K}}
\nc{\B}{{\cal B}}
\begin{center}

{\bf \Large  RG Equations and  High Energy Behaviour in \\[0.3cm]  Non-Renormalizable Theories} \vspace{1.0cm}

{\bf \large D. I. Kazakov$^{1,2}$} \vspace{0.5cm}

{\it
$^1$Bogoliubov Laboratory of Theoretical Physics, Joint
Institute for Nuclear Research, Dubna, Russia.\\
$^2$Moscow Institute of Physics and Technology, Dolgoprudny, Russia\\}
\vspace{0.5cm}

\abstract{We suggest a novel view on non-renormalizable interactions.  It is based on the usual  BPHZ $\R$-operation which is equally applicable  to any local QFT independently whether it is renormalizable or not.  As a playground we take the $\phi^4_D$ theory in $D$ dimensions for $D=4,6,8,10$ and consider the four-point scattering amplitude on shell. We derive the generalized RG equation and find the solution valid for any $D$ that sums up the leading  logarithms in all orders of PT  in full analogy  with the renormalizable case.  It is found that the scattering amplitude in the $\phi^4_D$  theory possesses the Landau pole at high energy for any D.  We discuss the application of the proposed procedure to other non-renormalizable theories.}
\end{center} 

\section{Introduction}

The Standard Model is based on renormalizable interactions. This was a matter of pecial concern and may serve as a selection criterion when looking for extensions of the SM. The reason is,
as is well known, that non-renormalizable interactions suffer from UV divergences and cannot be treated in a usual renormalization fashion since they require an infinite number of new types of counter terms. The other drawback is that  the amplitudes  in such theories increase with energy in each order of PT and one can not sum them up, like in renormalizable theories,  due to the absence of the proper formalism. 

Here we suggest a novel view on non-renormalizable theories, namely, we apply the renormalization procedure advocated in our earlier paper~\cite{PL}, where it was shown that one  can renormalize the theory in a usual way assuming that the renormalization constant $Z$ serves as an operator that depends on kinematics. This is the new and essential feature of the renormalization procedure which gives the UV finite amplitudes. Based on the BPHZ $\R$-operation, which is equally applicable in this case, one can  derive the  generalized RG equations for the scattering amplitude that sum up the leading divergences (asymptotics) in all orders of PT. After this, one can address the question of high energy behaviour of the amplitude. It is different for different theories just like in the renormalizable case and can be deduced from the one loop diagrams (in the renormalizable case it is the one loop beta-function).

In our previous papers~\cite{we}-\cite{we3} we chose as a playground for our analysis  the planar scattering amplitudes in D=8 super Yang-Mills theory considered within the spinor helicity formalism~\cite{Reviews_Ampl_General}. There were some reasons for that. Here we demonstrate the key points of our procedure on a simple example of  the scattering amplitudes in the $\phi^4_D$ theory in $D$-dimensions, where $D=4,6,8,10$. All these theories are treated in the same unified manner. It is shown that they all have similar UV behaviour possessing the Landau pole at high energy.

\section{Bogoliubov $\R$-operation and local counter terms}

Any local  QFT has the property that in higher orders of PT after subtraction of divergent subgraphs,
i.e.   after performing the incomplete ${\cal R}$-operation, the so-called ${\cal R}^\prime$-operation, the remaining UV divergences are local functions in the coordinate space or at maximum are polynomials of external momenta in momentum space. This follows from the rigorous proof of the Bogoliubov-Parasiuk-Hepp-Zimmermann $\R$-operation~\cite{BPHZ} and is equally valid in non-renormalizable theories as well. 

This property allows one to construct the so-called recurrence relations which relate the divergent contributions in all orders of perturbation theory (PT) with the lower order ones.  In renormalizable theories these relations are known as pole equations (within dimensional regularization) and are governed by the renormalization group \cite{tHooft}. The same is true, though technically is more complicated, in any local theory,  as we have demonstrated in \cite{we}-\cite{we3}.  We remind here some features of this procedure.

The incomplete ${\cal R}$-operation (${\cal R}^\prime$-operation) subtracts only the subdivergences of a given graph, while the full R operation is defined as 
\begin{equation}
\R G = (1-\K) \R' G,
\end{equation}
where ${\cal K}$ is an operator that singles out the singular part of the graph and $K{\cal R}^\prime G$- is the counter term corresponding to the graph G~\cite{Rop}.

After applying the ${\cal R}^\prime$-operation to a given graph in the n-th order of PT, one gets the following series of terms in dimensional regularization (we keep the leading poles only)
\beqa
{\cal R'}G_n&=&\frac{\A_n^{(n)}(\mu^2)^{n\epsilon}}{\epsilon^n}+\frac{\A_{n-1}^{(n)}(\mu^2)^{(n-1)\epsilon}}{\epsilon^n}+ ... +\frac{\A_1^{(n)}(\mu^2)^{\epsilon}}{\epsilon^n}\nonumber \\
&+&\mbox{lower\ pole\ terms} + \epsilon^0 \ \mbox{term},\label{Rn}
\eeqa
where the terms like $\frac{\A_{k}^{(n)}(\mu^2)^{k\epsilon}}{\epsilon^n}$  come from the $k$-loop graph which survives after subtraction of the $(n-k)$-loop counter term.
The resulting expression has to be local, hence does not contain terms like $\log^l{\mu^2}/\epsilon^k$, for any $l$ and $k$. This requirement leads to a sequence of relations for $\A_i^{(n)}$ which can be solved in favour of the lowest order term $\A_1^{(n)}$
\beqa
\A_n^{(n)}&=&(-1)^{n+1}\frac{\A_1^{(n)}}{n}.
\label{rel}
\eeqa
It is also useful to write down the local expression for the ${\cal KR'}$ terms (counter terms) equal to
\beq
{\cal KR'}G_n=\sum_{k=1}^n \left(\frac{\A_k^{(n)}}{\epsilon^n} \right)\equiv
\frac{\A_n^{(n)'}}{\epsilon^n}
\eeq
It is given by a similar expression
\beqa
\A_n^{(n)'}&=&(-1)^{n+1}\A_n^{(n)}=\frac{\A_1^{(n)}}{n}.
\label{rel2}
\eeqa
	
This means that, performing the  ${\cal R}'$-operation in order to extract the leading pole, one can only take care of the one loop  diagrams that survived after contraction and get the desired leading pole terms via eq.(\ref{rel}).  They can be calculated  in all loops pure algebraically starting from the one loop term $\A_1^{(1)}$. The same is true for subleading, subsubleading,  etc. poles  as well (see \cite{KV}), but one should take into account the diagrams with two, three, etc. loops, respectively, just like it takes place in renormalizable theories.   Here we restrict ourselves  to the leading  poles only.

To be more specific, let us consider as an example the scattering amplitudes in the $\phi^4_D$ theory in $D$-dimensions where $D=4,6,8,10$. (In what follows we take $D$ to be an integer while dimensional regularization is achieved by taking the integrals in $D-2\epsilon$ dimensions.) 

In what follows we consider the $2 \to 2$ scattering amplitude on shell and take the massless case. This means that all $p_i^2=0$ and the amplitude depends on the Mandelstam variables
$s,t,u$ with $s+t+u=0$. To get the full amplitude, one has to take into account all three channels. 

\begin{figure}[h]\vspace{0.3cm}
\begin{center}
\includegraphics[scale=0.45]{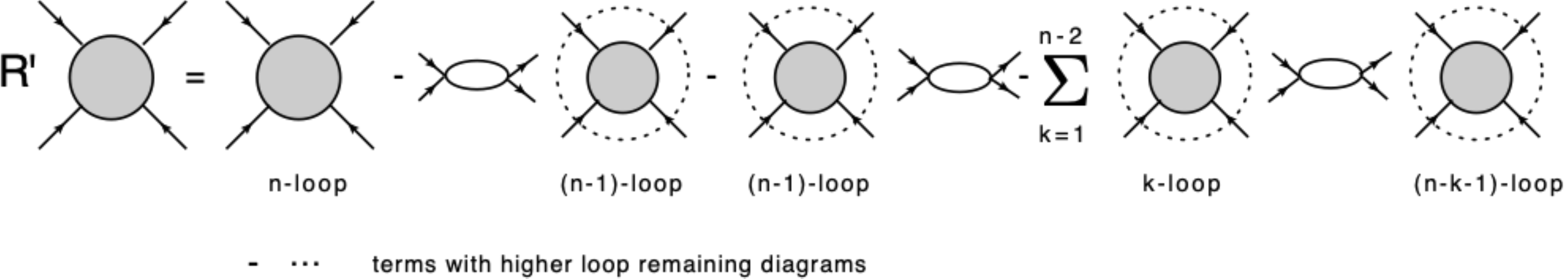}
\end{center}
\caption{The $\R'$-operation forу the 4-point function in the $\phi^4$ theory. To simplify the picture, we ignored here the difference between  the $s,t$ and $u$ channels.}\label{Rprime}
\end{figure}

\begin{figure}[h]
\begin{center}
\includegraphics[scale=0.43]{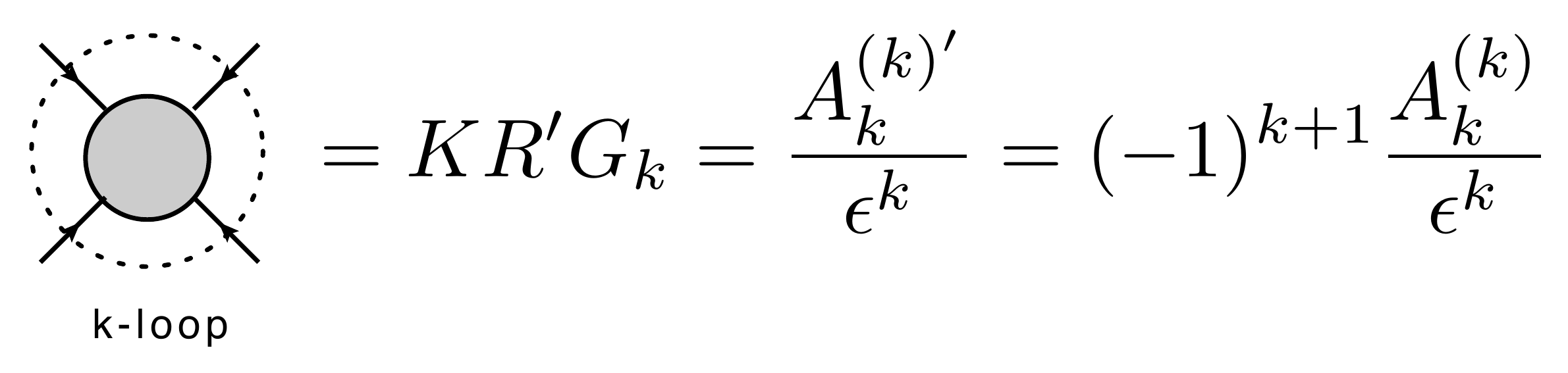}
\end{center}\vspace{-0.3cm}
\caption{The counter term $K\R'\ G$. The leading divergence is shown} \label{kr'}
\end{figure}

The $\R'$-operation for the 4-point function is shown schematically in Fig.\ref{Rprime}, where
the dotted line denotes the counter term obtained by the action of the $\R'$-operation on the corresponding subgraph (see Fig.\ref{kr'}). 


The  action of the $\R'$-operation shown in Fig.\ref{Rprime} is almost universal for any theory. The specific feature of the $\phi^4$ interaction is that the first two terms, which contain the live loop on the left or right,  are reduced to a bubble  diagram, while in general it might be a triangle as well.  The next term which contains the live loop in the middle is always a bubble. Remind also that only those diagrams give contribution to the $n$-th order pole in $n$ loops that contain divergent subgraphs starting from $1$ to $(n-1)$ loops.

\section{Recurrence relations and generalized RG equations}

We are now in a position to write down the recurrence relation for the leading poles on the basis of eq.(\ref{rel}) and Fig.\ref{Rprime}.  Define the four-point function $\Gamma_4$  as follows:
\beq
\Gamma_4(s,t,u)= \lambda  \bar \Gamma_4(s,t,u)=\lambda(1+\Gamma_s(s,t,u)+\Gamma_t(s,t,u)+\Gamma_u(s,t,u)),
\eeq
where the functions $\Gamma_t(s,t,u)$ and $\Gamma_u(s,t,u)$ are related to $\Gamma_s(s,t,u)$
by the cyclic change of the arguments.  $\Gamma_4$ obeys the PT loop expansion over $\lambda$
\beq
\Gamma_s=\sum_{n=1}^\infty (-z)^n S_n, \ \ \Gamma_t=\sum_{n=1}^\infty (-z)^n T_n, \ \ \Gamma_u=\sum_{n=1}^\infty (-z)^n U_n, \ \ \ z\equiv \frac \lambda\epsilon,
\eeq
where we keep only the leading pole terms.
To calculate them, we use the power of eq.(\ref{rel}). Indeed, in our notation $S_n+T_n+U_n=\A^{(n)}_n$ and can be expressed through $\A^{(n)}_1$ which is given by the diagrams shown in Fig.\ref{Rprime}. The dotted diagrams corresponding to $\K\R'G_k$ are given by $\A_k^{(k)'}$ and are also expressed through $\A^{(k)}_1$, according to eq.(\ref{rel2}).

The peculiarity of the procedure is that in the non-renormalizable case
the pole terms $S_n, T_n$ and $U_n$ depend on kinematics and are the polynomials over $s,t$ and $u$. This means that when substituting the counter terms into the recurrence relation, one has to integrate these kinematic factors over the remaining loop. This integration can be done in a general form leading to the following recurrence relation  for the s-channel part $S_n$ (and the same for $T_n$ and $U_n$ with the change of the arguments)

\beqa
&&n S_n(s,t,u)\nonumber\\
&=&\frac{s^{D/2-2}}{\Gamma(D/2-1)}\! \!\int_0^1 \!\! \! dx [x(1\!-\! x)]^{D/2-2}\left( S_{n-1}(s,t',u')\!+\! T_{n-1}(s,t',u')\!+\! U_{n-1}(s,t',u')\right)\nonumber\\
&+&\frac 12 \frac{s^{D/2-2}}{\Gamma(D/2-1)}\! \!\int_0^1 \!\! \! dx [x(1\!-\! x)]^{D/2-2}\sum_{k=1}^{n-2}\sum_{p=0}^{(D/2-2)k}\sum_{l=0}^{p}\frac{1}{p!(p+D/2-2)!}\times \label {rec}\\
&\times&\frac{d^p}{dt'^l du'^{p-l}}(S_k+T_k+U_k)\frac{d^p}{dt'^l du'^{p-l}}(S_{n-k-1}+T_{n-k-1}+U_{n-k-1})s^p[x(1-x)]^p t^l u^{p-l}\nonumber
\eeqa
with $t'=- xs, u'=-(1-x)s$. The first linear term of eq.(\ref{rec}) corresponds to the first two diagrams in Fig.\ref{Rprime} and the  second nonlinear term is due to the third diagram  with the  live loop in the middle. Integration over $x$ is just integration over the Feynman parameter in the loop diagram. Multiple sums appear due to the $g_{\mu\nu}$ factors aising when integrating multiple momenta in the numerator of the diagrams.

This recurrence relation allows one to calculate all the leading divergences in all loops in a pure algebraic way starting from the one loop diagram. Thus, taking the one loop values
\beq
S_1=\frac 12 \frac{\Gamma(D/2-1)}{\Gamma(D-2)}s^{D/2-2}, \  \ T_1=\frac 12 \frac{\Gamma(D/2-1)}{\Gamma(D-2)}t^{D/2-2}, \  \ U_1=\frac 12 \frac{\Gamma(D/2-1)}{\Gamma(D-2)}u^{D/2-2}  \label{1}, 
\eeq
one immediately gets from (\ref{rec})
\beq
S_2=\frac 14  \frac{\Gamma(D/2-1)}{\Gamma(D-2)}\left[  \frac{\Gamma(D/2-1)}{\Gamma(D-2)}+2(-)^{D/2} \frac{\Gamma(D-3)}{\Gamma(3D/2-4)}\right]s^{D-4}, \ \ \ etc.  \label{2}
\eeq

One can convert the recurrence relation (\ref{rec}) into a differential equation for the function $\Gamma_s(s,t,u|z)$ taking the sum over $n$ of eq.(\ref{rec}). Thus, taking the sum  $\sum_{n=2}^\infty (-z)^{n-1} $, one gets
\beqa
&-&\frac{d \Gamma_s(s,t,u)}{dz}=\frac 12 \frac{\Gamma(D/2-1)}{\Gamma(D-2)}s^{D/2-2}\nonumber \\ &+&\frac{s^{D/2-2}}{\Gamma(D/2-1)}\! \!\int_0^1 \!\! \! dx [x(1\!-\! x)]^{D/2-2}\left[\Gamma_s(s,t',u')\!+\! \Gamma_t(s,t',u')\!+\! \Gamma_u(s,t',u')\right]\vert_{\scriptsize  \begin{array}{l}
t'=- xs, \\ u'=-(1-x)s\end{array}}\nonumber\\
&+&\frac 12 \frac{s^{D/2-2}}{\Gamma(D/2-1)}\! \!\int_0^1 \!\! \! dx [x(1\!-\! x)]^{D/2-2}\sum_{p=0}^{\infty}\sum_{l=0}^{p}\frac{1}{p!(p+D/2-2)!}\times \label {eqs}\\
&\times&\left(\frac{d^p}{dt'^l du'^{p-l}}(\Gamma_s+\Gamma_t+\Gamma_u)\vert_{\scriptsize  \begin{array}{l}
t'=- xs, \\ u'=-(1-x)s\end{array}}\right)^2 s^p[x(1-x)]^p t^l u^{p-l}, \nonumber
\eeqa
with the boundary condition $\Gamma_s(z=0)=0$
and the same for $\Gamma_t$ and $\Gamma_u$ with the change of the arguments. 

Equation (\ref{eqs}) can be simplified if written for the whole function $\bar \Gamma_4$.  One can also notice that due to the one-to-one correspondence between $1/\epsilon$ and $\log\mu^2$  one can rewrite equation for $\bar \Gamma_4$ in a more familiar way
\beqa
&&\frac{d \Gamma_s(s,t,u)}{d\log\mu^2}=
-\frac{\lambda}{2} \frac{s^{D/2-2}}{\Gamma(D/2-1)}\! \!\int_0^1 \!\! \! dx [x(1\!-\! x)]^{D/2-2}\sum_{p=0}^{\infty}\sum_{l=0}^{p}\frac{1}{p!(p+D/2-2)!}\times \nonumber\\
&&\times \left(\frac{d^p \bar \Gamma_4(s,t',u')}{dt'^l du'^{p-l}}\vert_{\scriptsize  \begin{array}{l}
t'=- xs, \\ u'=-(1-x)s\end{array}}\right)^2 s^p[x(1-x)]^p t^l u^{p-l},
 \label {eqrg}
\eeqa
with the boundary condition $\Gamma_s(\log\mu^2=0)=0$. The same equation is also valid for $\Gamma_t$  and $\Gamma_u$ with the cyclic change of the arguments. 

Equation (\ref{eqrg}) is nothing more than the desired generalized RG equation for the $\phi^4_D$ theory in $D$-dimensions. To see it, consider the case when $D=4$. It corresponds to a well-known renormalizable theory where the amplitude $\bar \Gamma_4$ does not depend on $s,t$ and $u$ and hence one can drop the integrals and sums  in eq.(\ref{eqrg}). Adding the terms with $\Gamma_s, \Gamma_t$ and $\Gamma_u$ together, one has
\beq
D=4: \ \ \ \ \ \ \  \frac{d\bar \Gamma_4}{d\log\mu^2}=
-\frac 32\lambda \bar \Gamma_4^2, \ \ \ \ \bar \Gamma_4(\log\mu^2=0)=1. \label{d4}
\eeq

\section{High energy behaviour of the $\phi^4_D$ theory}

To find the high energy behaviour of the amplitude $\Gamma_4$ when $s\sim t\sim u\sim E^2$, one has to solve eq.(\ref{eqrg}).   In the case of $D=4$, eq.(\ref{d4}) has
 an obvious solution in the form of a geometrical progression
\beq
\bar \Gamma_4=\frac{1}{1+\frac 32\lambda \log(\mu^2/E^2)}\ \ \ \ \ or\ \ \ \ \    \Gamma_4=\frac{\lambda}{1+\frac 32\lambda \log(\mu^2/E^2)}.
\eeq

This solution suggests the form of the solution to eq.(\ref{eqrg}) for arbitrary $D$.  We dare say it is
\beq
\Gamma_4(s,t,u)={\mathcal P} \frac{\lambda}{1+\frac 12 \frac{\Gamma(D/2-1)}{\Gamma(D-2)}\lambda (s^{D/2-2}+t^{D/2-2}+u^{D/2-2})\log(\mu^2/E^2)},
\label{sol}
\eeq
where the symbol ${\mathcal P}$ means the ordering in a sense of eq.(\ref{rec}), i.e.  when expanding the geometrical progression in a series over $\lambda$, one has to choose a single loop in the $s,t$ or $u$ channel and then integrate the powers of $s,t$ and $u$ over this loop. This gives exactly the PT series of the form (\ref{rec}).  Symbolically, one can write eq.(\ref{sol}) as
 \beq
\Gamma_4(s,t,u)={\mathcal P} \frac{\lambda}{1+\lambda A_1^{(1)}\log(\mu^2/E^2)}
\label{sol2}.
\eeq
Perturbative expansion then looks like
\beq
{\mathcal P} \sum_{n=0}^{\infty} (-\lambda)^n \log^n(\mu^2/E^2) (A_1^{(1)})^n
\eeq
where the n-th term  has to be understood as
\beq
{\mathcal P}(A_1^{(1)})^n = \int_0^1 dx \sum_{k=0}^{n-1}\  \overrightarrow{ {\mathcal P}(A_1^{(1)})^k }\ A_1^{(1)} \ \overleftarrow{{\mathcal P}(A_1^{(1)})^{n-1-k}}, 
\eeq
where the arrow means that one has to integrate the expression under the arrow sign through 
$A_1^{(1)} $ in a sense of eq.(\ref{rec}).

 Consider how it works in the lowest orders of PT. Expanding the  geometrical progression (\ref{sol}) over $\lambda$ in the second order, one gets in the $s$ channel
 \beq
\sim \lambda^2 s^{D/2-2}(s^{D/2-2}+t^{D/2-2}+u^{D/2-2})
 \eeq
and similar for the $t$ and $u$ channels. Now taking  the $s$ channel, one has to integrate the bracket over the one loop s-channel diagram which gives exactly the first line of eq.(\ref{rec}) with the resulting expression for $S_2$ (\ref{2}).  In the third order one has
\beq
\sim \lambda^3 s^{D/2-2}(s^{D/2-2}+t^{D/2-2}+u^{D/2-2})^2
 \eeq
 and besides the first line of eq.(\ref{rec}) one also has contributions where the $t$ and $u$ terms come from the diagrams standing to the left and right of the s-channel one.  This leads to the contributions given by the second line of eq.({\ref{rec}).

 Thus, expanding the geometrical progression (\ref{sol}) with the proper ordering, we reproduce the whole PT series of the leading poles (logarithms).  
At the same time, the  solution (\ref{sol}) gives us an excellent opportunity to study high energy behaviour of the whole function $\Gamma_4$. One can look for singularities of  the solution (\ref{sol}) under the sign of ${\mathcal P}$-ordering and make conclusions about the  high energy behaviour of the amplitude.  Consider some particular cases:

D=4: One obviously has a Landau pole when $E\to \infty$ due to a positivity of the coefficient 3/2, as it should be.

D=6: One has $s+t+u=0$ in the denominator. This means that the leading divergences (logarithms) cancel. One can be convinced that this is indeed the case considering eq.(\ref{rec}) or explicitly check that $S_2$ given by eq.(\ref{2})  equals zero.

D=8: One has $s^2+t^2+u^2 > 0$, i.e. one again has a Landau pole when $E\to \infty$ but it appears much quicker due to the power law behaviour.

D=10: One has $s^3+t^3+u^3=3\ s\ t\ u>0$ since $s>0$ and $t,u<0$. Hence, one again has a Landau pole.

Thus, we come to a conclusion that in the $\phi^4_D$ theory in arbitrary $D$ one has the Landau pole behaviour at high energies as it is the case with renormalizable interactions. This is defined by the sign of the one loop diagram given by eq.(\ref{1}) for all $D$.

\section{Discussion}
Apparently, the reasoning above is not specific to the $\phi^4_D$ interaction and can be repeated in any theory. To get the desired recurrence relation, which is the starting point of our analysis, one has to write down the $\R$-operation and analyze the one loop diagrams left after subtraction of the counter terms. In the case of the $\phi^4$ theory one always has a bubble, while in the maximally SUSY gauge theories one has a triangle~\cite{we}.  The difference is not principle,  the recurrence relations look similar.  At the same time, as is well known, the sign of the one loop diagram in gauge theories is negative. This opens up the possibility to get  an asymptotically free theory in higher dimensions. It does not happen in the maximally supersymmetric theories due to cancellation of  bubble diagrams but may occur in the usual Yang-Mills theories. 

It should be stressed that the structure of the recurrence relation and the corresponding RG equation reflects the  structure of the $\R$-operation and is universal.  Limited to only leading divergences it is reduced to the  one loop diagrams and,  as it follows from Fig.\ref{Rprime}, contains only the linear and quadratic terms. This fact makes it possible to conduct a general analysis of the solutions and
asymptotical regimes as one has in renormalizable theories. 

\section*{Acknowlegments}
The author is grateful to A.Borlakov abd D.Tolkachev  for useful discussions and check of the calculations of the diagrams. Special thanks to S.Mikhailov for reading the manuscript and useful comments. This work was supported by the Russian Science Foundation grant \# 16-12-10306.

\end{document}